\documentclass[3p,times]{elsarticle}

\usepackage{ecrc}

\volume{00}

\firstpage{1}

\journalname{Nuclear Physics A}

\runauth{}

\jid{npa}

\jnltitlelogo{Nuclear Physics A}

\usepackage{amssymb}

\biboptions{square,comma,numbers,sort&compress}

\usepackage[figuresright]{rotating}

\begin{document}

\begin{frontmatter}



\dochead{}

\title{Jet quenching effects on the anisotropic flow at RHIC}


\author{R.~P.~G.~Andrade,$^1$ J.~Noronha,$^2$ Gabriel S. Denicol$^3$}

\address{$^1$Centro de Ci\^{e}ncias Naturais e Humanas, Universidade Federal do ABC, Santo Andr\'{e}, SP, 09210-170, Brazil\\
$^2$Instituto de F\'{\i}sica, Universidade de S\~{a}o Paulo, C.P. 66318,05315-970 S\~{a}o Paulo, SP, Brazil\\
$^3$Department of Physics, McGill University, 3600 University Street, Montreal, Quebec, H3A, 2T8, Canada}

\begin{abstract}
In this article we investigate how the energy and momentum deposited by partonic dijets in the quark-gluon plasma may perturb the geometry-induced hydrodynamic expansion of the bulk nuclear matter created in heavy ion collisions at RHIC. The coupling between the jets and the medium is done through a source term in the energy-momentum conservation equations for ideal hydrodynamics. We concentrate our attention at mid-rapidity and solve the equations event-by-event imposing boost-invariance. For $p_T \gtrsim 1$ GeV the anisotropic flow is found to be considerably enhanced, if the dijets deposit on average more than 12 GeV in the medium (or equivalently 6 GeV for each jet of the pair), which corresponds, in our model, to an average suppression greater than $65\%$ of the initial jet transverse energy.
\end{abstract}

\begin{keyword}


Relativistic hydrodynamics; Jet quenching; Fourier coefficients of the flow
\end{keyword}

\end{frontmatter}


\section{Introduction}
\label{sec:introduction}

The jet suppression observed in relativistic heavy ion collisions performed at the Relativistic Heavy Ion Collider (RHIC) and the Large Hadron Collider (LHC) \cite{Adams:2003kv,Adcox:2001jp,Adams:2003im,Adler:2003ii,CMS:2012aa,Abelev:2012hxa,Aamodt:2010jd} is a strong evidence that a hot and dense nuclear matter, usually called quark gluon plasma (QGP), is created in these experiments. The observation of such a phenomenon ushered in an ample scenario for the study of how the jet spectra are modified by the interaction with the medium (see, for instance, \cite{Burke:2013yra} and references therein). However, the counterpart, i.e., the study of how the medium is modified by the interaction with the jets, has been poorly explored. The aim of this article is to improve our knowledge of the effects of the dijets on the hydrodynamic evolution of the QGP. Using a 2+1 hydrodynamic model on an event-by-event basis, we try to understand such effects through of the anisotropic flow parameters $\left\{v_n,\Psi_n \right\}$, namely the $n$th Fourier coefficient of the azimuthal distribution of hadrons and the respective phase. Naturally, depending on the amount of energy-momentum deposited in the medium by the dijets, the jet-induced anisotropic flow may represent a non-negligible fraction of the total anisotropic flow. All the results presented in this article correspond to Au+Au collisions at 200A GeV in the $(0-5)\%$ centrality window. We use hyperbolic coordinates, i.e., $\tau=\sqrt{t^2-z^2}$, $\eta=0.5\ln \left[ \left(t+z \right)/\left(t-z \right)  \right]$ and $\vec{r}=\left(x,y \right)$. The initial time at which we begin the hydrodynamic evolution $\tau_0=1$ fm. In addition, $\hbar=k_{B}=c=1$.

\section{Hydrodynamic model}
\label{sec:hydro}

Based on the assumption that the energy-momentum lost by the jets quickly thermalizes in the QGP \cite{Chaudhuri:2005vc}, the coupling between the dijets and the medium, in our model, is done through a source term in the energy-momentum conservation equations. Thus, in the ideal fluid approximation, one finds that

\begin{equation}
D_{\mu}T^{\mu \nu}=J^{\nu},
\label{eq:em_cons}
\end{equation}

\noindent
where $T^{\mu \nu}=\omega u^{\mu} u^{\nu} - pg^{\mu \nu}$ is the ideal fluid energy-momentum tensor, $\omega$ is the enthalpy, $p$ the pressure, $u^\mu$ the fluid 4-velocity, $D_\mu$ the covariant derivative, and $g^{\mu \nu}$ is the metric tensor. In our model, this equation is solved assuming boost-invariance. The 4-current density $J^\nu$ (the source) is parameterized, in the laboratory frame, as (see, for instance, Refs.\ \cite{Chaudhuri:2005vc,Betz:2010qh,Betz:2009su,Torrieri:2009mv,Betz:2008ka,Betz:2008wy,Andrade:2014swa})

\begin{equation}
J^{\nu}\left(\tau, \vec{r}\right)= \sum_{n=1}^{n_p}    \frac{s \left(\vec{r}^{\hspace{0.5mm} jet}_n \left(\tau \right) \right)}{s_0} \left. \frac{dE}{d l} \right|_0         F \left(\vec{r} - \vec{r}^{\hspace{0.5mm} jet}_n \left(\tau \right),\tau; \sigma \right)       \left(1,\vec{v}^{\hspace{0.5mm} jet}_n,0 \right),
\label{eq:current1}
\end{equation}

\noindent
where $n_p$ is the number of partonic jets, $\vec{r}^{\hspace{0.5mm} jet}_n$ and $\vec{v}^{\hspace{0.5mm} jet}_n$ (with $|\vec{v}^{\hspace{0.5mm} jet}_n |=1$) are the position and velocity of the $n$th parton that moves in a straight line on the mid-rapidity transverse plane, $s (\vec{r}^{\hspace{0.5mm} jet}_n )$ is the entropy density computed at the position of the $n$th parton, and the function $F$ \cite{Chaudhuri:2005vc,Andrade:2014swa} corresponds to a Gaussian shaped source of width $\sigma$ (we set $\sigma=0.6$ fm). We assume only one dijet per event, consequently, $n_p=2$,  $\vec{r}^{\hspace{0.5mm} jet}_{1} (\tau_0) =\vec{r}^{\hspace{0.5mm} jet}_{2} (\tau_0)$ and $\vec{v}^{\hspace{0.5mm} jet}_1=-\vec{v}^{\hspace{0.5mm} jet}_2$. The parameters $\left. dE/d l \right|_0$ and $s_0$ are the reference energy loss rate ($l=\tau - \tau_0$ is transverse distance traveled by the partons) and the reference entropy density. The latter corresponds to the maximum of the average entropy density distribution in the $(0-5)\%$ centrality window (in our model, $s_0=70$ fm$^{-3}$). The former is a free parameter, which is varied from 5 to 20 GeV/fm. Naturally, the bigger this parameter is, the more energy-momentum the jets deposit in the medium.

We use the equation of state EOS S95n-v1 \cite{Huovinen:2009yb}, which combines results from lattice QCD at high temperatures and the hadron resonance gas equation at low temperatures. To compute the particle spectrum, we use the Cooper-Frye prescription \cite{Cooper:1974mv}. In this method, the particles escape from the fluid after crossing a hyper-surface of constant temperature, usually called freeze-out temperature, $T_{fo}$ (we set $T_{fo}=0.14$ GeV). Finally, all the results presented in this paper correspond to positively charged pions directly emitted from the freeze-out hyper-surface.

In few words, the procedure to compute an observable event-by-event, including the jet parametrization, is the following: (i) the initial conditions for hydrodynamics are computed using an implementation of the Monte Carlo Glauber model \cite{Drescher:2006ca,Drescher:2007ax}; (ii) the initial position of the dijet (one dijet per event) is chosen on the mid-rapidity transverse plane (see the details in Ref. \cite{Andrade:2014swa}); (iii) the dijet azimuthal angle in chosen isotropically; (iv) the initial jet transverse energy $ E_{T}^{\hspace{0.5mm} jet}$ (the same for both jets in the pair) is chosen according to the jet yield per event in p+p collisions scaled by the number of binary collisions in Au+Au collisions \cite{Andrade:2014swa,Salur:2008hs}; (v) the hydrodynamic evolution is computed through the SPH method \cite{Aguiar:2000hw,Andrade:2013poa} and (vi) the final spectra (for direct positively charged pions) is computed using the Cooper-Frye prescription \cite{Cooper:1974mv}. At the end of the simulation, the average value of a given observable is calculated over an ensemble of events. We define ``mixed ensemble" as the ensemble composed by 750 events without dijets and 250 with dijets (totaling 1000 events). This proportion is fixed by the jet yield per event \cite{Andrade:2014swa,Salur:2008hs}. On the other side, the ``jet ensemble" corresponds only to events with dijets (250 events).

\section{Results}
\label{sec:results}

\begin{figure}[ht]
\begin{center}
\includegraphics[scale=0.33]{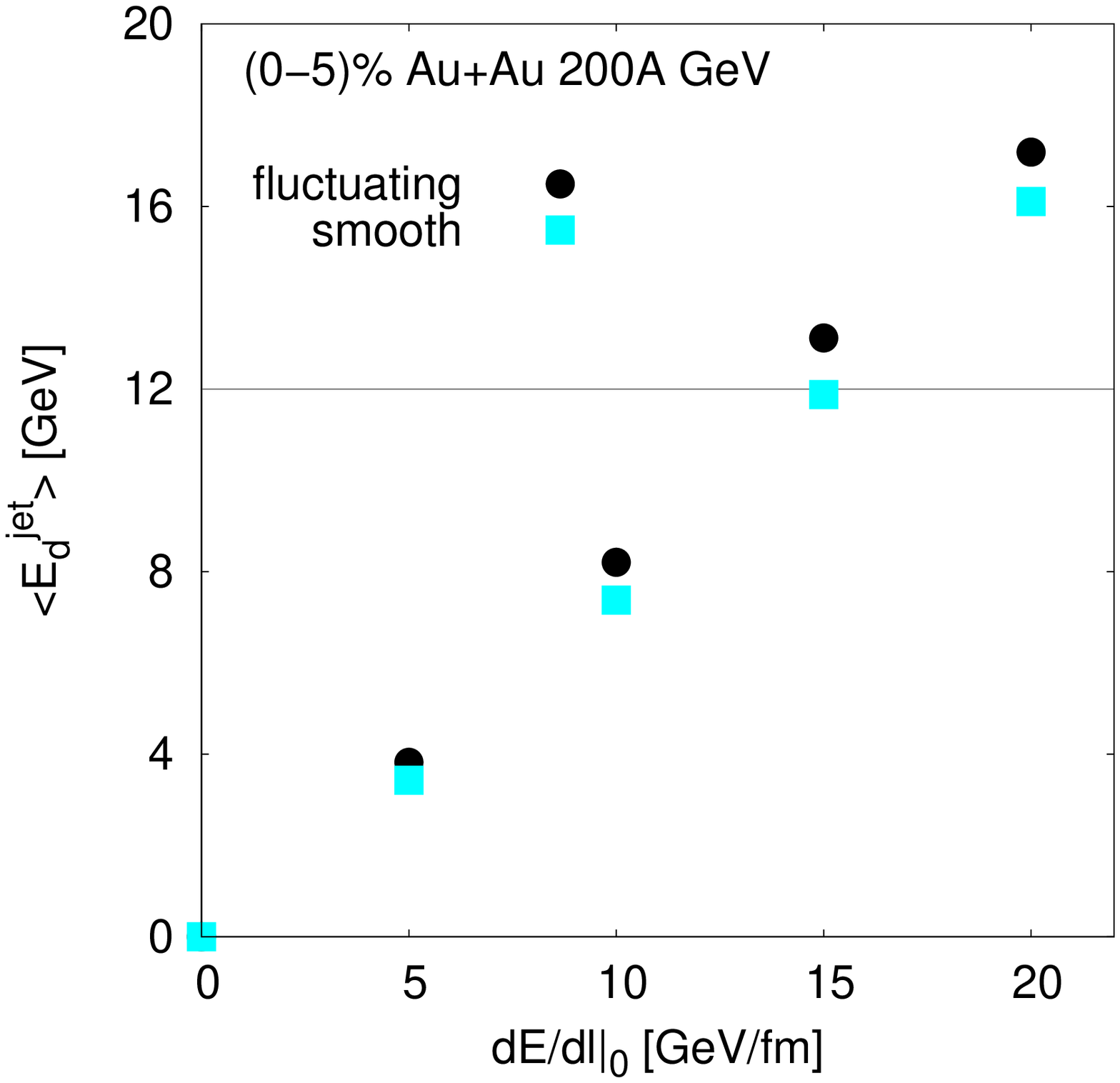}
\includegraphics[scale=0.33]{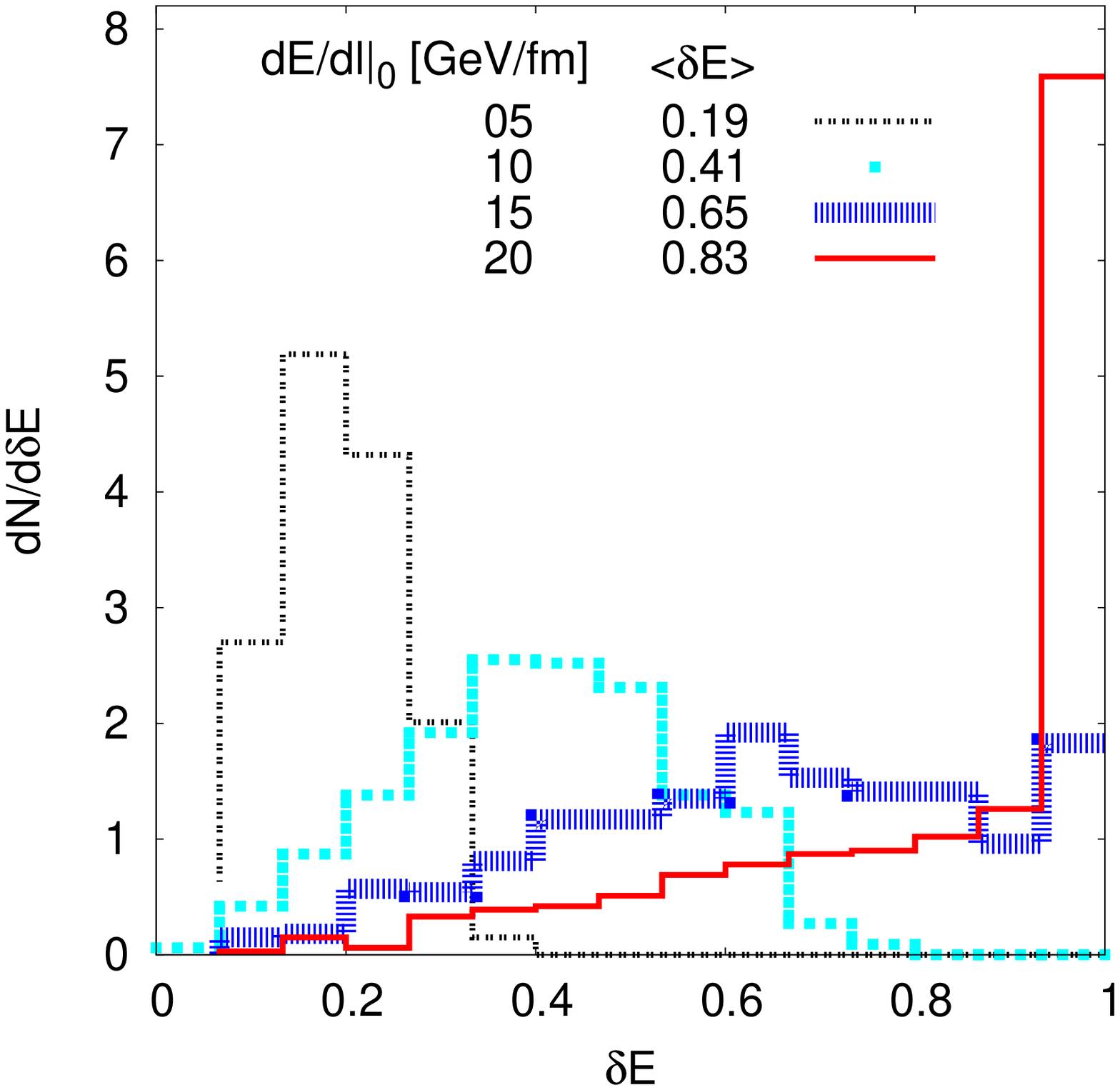}
\caption{\label{fig:jet_ael} (Color online) Left: average energy deposited in the medium, around mid-rapidity, by the dijet, $< E_{d}^{\hspace{0.5mm} jet}>$, in the $(0-5)\%$ centrality window, as a function of the reference energy loss rate $\left. dE/d l \right|_0$ (using the jet ensemble). Right: distribution of the ratio $\delta E = E_{d}^{\hspace{0.5mm} jet}/ E_{T}^{\hspace{0.5mm} jet}$ for four values of the parameter $\left. dE/d l \right|_0$. The respective average value $<\delta E>$ is shown on the plot.}
\end{center}
\end{figure}

In Fig.\ \ref{fig:jet_ael} (left) we show the average energy deposited in the medium by the dijet, $< E_{d}^{\hspace{0.5mm} jet}>$, in the $(0-5)\%$ centrality window, as a function of the reference energy loss rate $\left. dE/d l \right|_0$. To compute the curve labeled ``smooth" (squares), the fluctuating initial energy density distribution was replaced, in each event, by a smooth one while keeping unchanged the initial position of the dijet. As one can see, the fluctuations slightly enhance the suppression of jets in the medium. In the same figure (right), we show the distribution of the ratio $\delta E = E_{d}^{\hspace{0.5mm} jet}/ E_{T}^{\hspace{0.5mm} jet}$, i.e., the relative amount of energy (with respect to the initial jet transverse energy $ E_{T}^{\hspace{0.5mm} jet}$) that is lost to the medium, for four values of the parameter $\left. dE/d l \right|_0$. The respective average value $<\delta E>$ is shown on the plot. Observe that $<\delta E>$ gets close to unity when $\left. dE/d l \right|_0$ is increased. In fact, depending on the magnitude of the coupling between the jets and the QGP,  a considerable fraction of the jets may be completely absorbed by the medium. These distributions survey, in our model, an estimative of the suppression of the jets in the medium and can be used to calibrate the free parameter $\left. dE/d l \right|_0$. As we are going to see in the next plot, for $\left. dE/d l \right|_0 \gtrsim 15$ GeV/fm, which corresponds to a suppression on average greater than $65\%$,  the jet quenching effect may create relevant additional anisotropic flow.

\begin{figure}[ht]
\begin{center}
\includegraphics[scale=0.40]{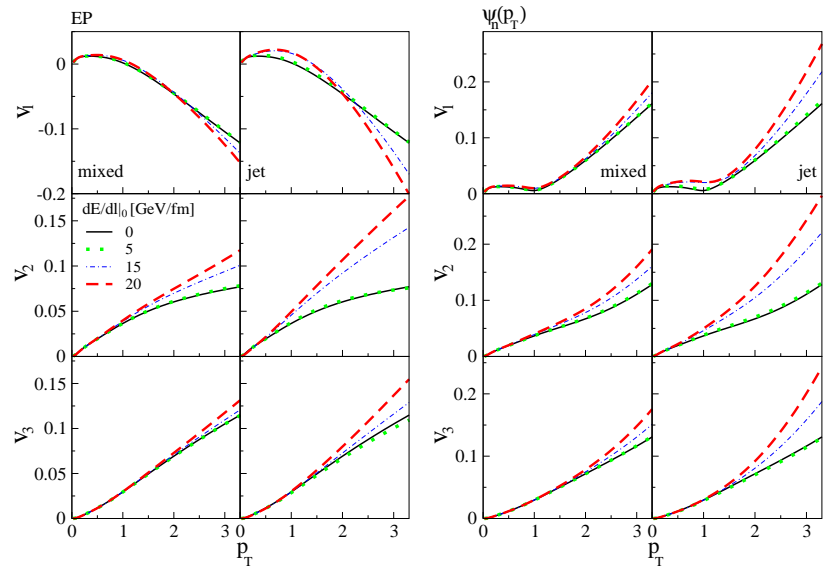}
\caption{\label{fig:vn_3} (Color online) Transverse momentum dependence of the $v_n$ coefficients ($n=1,2,3$) for four values of the parameter $\left. dE/d l \right|_0$. The left panels correspond to the event plane method. On the right panels we show the same observables computed using $\Psi_n=\Psi_n \left( p_T\right)$, i.e., the phase is computed for each $p_T$ bin \cite{Andrade:2013poa}. The panels labeled ``mixed" correspond to an ensemble of 1000 events that includes 750 events without and 250 events with dijets. The panels labeled ``jet" correspond to an ensemble of 250 events that includes only events with dijets.}
\end{center}
\end{figure}

In Fig.\ \ref{fig:vn_3}, we show the $v_n$ coefficients ($n=1,2,3$), as a function of the transverse momentum, for four values of the parameter $\left. dE/d l \right|_0$. The left panels correspond to the Event Plane method (EP) where the phase $\Psi_n$ is computed using all the hadrons of the event \cite{Poskanzer:1998yz}. The right panels correspond to method used in Ref. \cite{Andrade:2013poa} where $\Psi_n=\Psi_n \left( p_T\right)$, i.e., the phase is computed for each $p_T$ bin. Naturally, the latter procedure maximizes the anisotropy. The negative sign observed in the coefficient $v_1\left(p_T \right)$, computed using the event plane method, is a consequence of momentum conservation: if the low $p_T$ particles move in one direction, the higher $p_T$ particles must move in the opposite direction to conserve momentum. Note that, in the majority of the cases, for $\left. dE/d l \right|_0 > 5$ GeV/fm, the effects of the jets are important in the region of intermediate $p_T$ ($1 \lesssim p_T \lesssim 3$ GeV). In the region of low $p_T$ ($ p_T < 1$ GeV) the effects are negligible. On the other side, using our lower limit for the coupling between the QGP and the jets, $\left. dE/d l \right|_0=5$ GeV/fm, the results are nearly identical to the results without jets. Finally, the anisotropy is enhanced, as expected, when only events with dijets are included (the jet ensemble).

\section{Conclusions}
\label{sec:conclusion}
In our simplified model we found that the effects of the dijets on the hydrodynamic evolution of the QGP are less important in the region of low $p_T$ ($p_T < 1$ GeV). Even for the highest value of energy loss used in this paper, $dE/dl|_{0} =20$ GeV/fm, which corresponds to an average suppression of $83\%$ of the initial jet transverse energy, we found that the dijets affect mainly the region of intermediate $p_T$ ($1 \lesssim p_T \lesssim 3$ GeV). For a more detailed analysis see \cite{Andrade:2014swa}.

R.~P.~G.~Andrade and J.~Noronha thank
Funda\c c\~ao de Amparo \`{a} Pesquisa do Estado de S\~{a}o Paulo
(FAPESP) and Conselho Nacional de Desenvolvimento Cient\'{\i}fico e Tecnol%
\'{o}gico (CNPq) for financial support. G. S. Denicol acknowledges
the support of a Banting fellowship provided by the Natural Sciences and Engineering Research Council of Canada.





\bibliographystyle{elsarticle-num}
\bibliography{draft}







\end{document}